\documentclass[preprint,11pt]{aastex}
\usepackage{emulateapj5}
\newcommand{\ee}[1]{\mbox{${} \times 10^{#1}$}}
\newcommand{\eten}[1]{\mbox{$10^{#1}$}}

\newcommand\cmv{\mbox{cm$^{-3}$}}

\newcommand{\ISO}{\mbox{\it ISO}}
\newcommand{\mm}{millimeter}
\newcommand\submm{submillimeter}

\newcommand\fir{far-infrared}
\newcommand\mir{mid-infrared}

\newcommand\uv{ultraviolet}

\newcommand{\msun}{\mbox{M$_\odot$}}

\newcommand{\tk}{\mbox{$T_K$}}
\newcommand{\td}{\mbox{$T_d$}}

\newcommand{\meandev}{\mbox{$\langle \delta \rangle$}} 
\newcommand{\mean}[1]{\mbox{$\langle#1\rangle$}} 
\newcommand{\isrf}{\mbox{\rm{ISRF}}}

\newcommand{\hh}{\mbox{{\rm H}$_2$}}

\input{epsf}

\def\plotfiddle#1#2#3#4#5#6#7{\centering \leavevmode
\vbox to#2{\rule{0pt}{#2}}
\includegraphics{#1}}

\newcommand{\Snu}{\mbox{$S_{\nu}$}}

\newcommand{\nInu}{\mbox{$I^{norm}(b)$}}
\newcommand{\nInum}{\mbox{$I^{norm}_{mod}(b)$}}
\newcommand{\kappanu}{\mbox{$\kappa_{\nu}$}}
\newcommand{\Td}{\mbox{$T_{d}$}}
\newcommand{\Tk}{\mbox{$T_{K}$}}
\newcommand{\Tdr}{\mbox{$T_{d}(r)$}}
\newcommand{\Tkr}{\mbox{$T_{K}(r)$}}

\newcommand{\ppc}{pre-protostellar core}

\newcommand{\nc}{\mbox{$n_c$}}
\newcommand{\chisq}{\mbox{$\chi_r^2$}}

\newcommand{\router}{\mbox{$r_o$}}
\newcommand{\rinner}{\mbox{$r_i$}}
\newcommand{\rflat}{\mbox{$r_{flat}$}}

\slugcomment{\footnotesize {}}
\shorttitle{Modeling Pre-Protostellar Cores}
\shortauthors{Evans et al. }

\begin{document}


\title {\bf Tracing the Mass during Low-Mass Star Formation. 
II. Modelling the Submillimeter Emission from Pre-Protostellar Cores}
\author {Neal J. Evans II\altaffilmark{1}\altaffilmark{2} }
\affil{Department of Astronomy, The University of Texas at Austin,
       Austin, Texas 78712--1083}
\altaffiltext{1}{University College London, Gower Street, London WC1E 6BT, UK}
\altaffiltext{2}{Leiden Observatory, P. O. Box 9513, 2300 RA Leiden, the
Netherlands}
\email{nje@astro.as.utexas.edu}
\author{Jonathan M. C. Rawlings}
\affil{Department of Physics and Astronomy, University College London, Gower
Street, London WC1E 6BT, UK}
\email{jcr@star.ucl.ac.uk}
\author{Yancy L. Shirley}
\affil{Department of Astronomy, The University of Texas at Austin,
       Austin, Texas 78712--1083}
\email{yshirley@astro.as.utexas.edu}
\author{Lee G. Mundy}
\affil{Department of Astronomy, University of Maryland, College Park, MD}
\email{lgm@astro.umd.edu}

\begin{abstract}

We have modeled the emission from dust in pre-protostellar cores, including a
self-consistent calculation of the temperature distribution
for each input density distribution. Model density distributions include 
Bonnor-Ebert spheres and power laws. The Bonnor-Ebert
spheres fit the data well for all three cores we have modeled. 
The dust temperatures
decline to very low values ($\Td \sim 7$ K) in the centers of
these cores, strongly affecting the dust emission. Compared to
earlier models that assume constant dust temperatures, our models indicate
higher central densities and smaller regions of relatively constant
density. Indeed, for L1544, a power-law density distribution, similar to that
of a singular, isothermal sphere, cannot be ruled out.
For the three sources modeled herein, there seems to be a sequence of
increasing central condensation, from L1512 to L1689B to L1544.
The two denser cores, L1689B and L1544, 
have spectroscopic evidence for contraction, suggesting
an evolutionary sequence for pre-protostellar cores.

\end{abstract}


\section{Introduction}

Molecular cloud cores that have not yet formed a star provide the
best opportunity to determine the initial conditions for star formation.
The density distribution as a function of radius in such starless cores
is a strong discriminator between theoretical models, and it provides an
essential tool for understanding the process of collapse and star formation
that follows \citep{Andre et al. (2000)}. 
However, the earliest evolution of a cloud core is very poorly
constrained, and its evolutionary track is unclear.

For those more evolved objects that possess infrared sources, the sequence is 
well described by the empirical evolutionary sequence of \citet{Lada (1987)}, 
which is supported by theoretical modeling of the spectral energy
distribution (SED) \citep{Adams et al. 
(1987)}. In this scheme, the well-known classification (i.e.
Class I, II and III) is based on the near-infrared spectral index and
is interpreted in terms of the sources becoming progressively less embedded in
circumstellar dust.
The scheme was extended to younger, more deeply embedded objects with the
Class 0 classification of \citet{Andre et al. (1993)}. These objects possess a
highly enshrouded IRAS source (so that their SEDs peak longward of 100 
\micron); accretion models suggest that the mass of the central object is less
than  or similar to that of the collapsing envelope.
Once a protostellar core has formed, the density distributions are 
strongly centrally peaked and the evolution of the objects has a ``singular''
nature in that the dynamics are primarily determined by the characteristics of
the innermost regions of the cores. 

At earlier stages of evolution, starless cores, defined as
dense cores of gas that do not possess an IRAS source \citep{Myers and 
Benson (1983), Benson and Myers (1989)}, are potential sites of 
star-formation.
An important breakthrough in the study of these objects was the detection of
dust continuum emission at \mm\ and \submm\ wavelengths by 
\citet{Ward-Thompson et al. (1994)}, who dubbed the subset of starless cores 
that possess such emission ``pre-protostellar cores'' (hereafter PPCs).

However, unlike the Class 0--III sources described above, the classification of
a source as a PPC can have a much broader interpretation:
the density distribution in these objects could be anything between 
uniform and strongly centrally peaked. They may or may not be gravitationally
bound. Moreover, the dynamical and evolutionary status of PPCs is poorly
defined; they may be quasi-statically contracting (perhaps supported by
non-thermal  pressures), or in an early stage of dynamic collapse. 
Alternatively, it
is possible  that a combination of magnetic, turbulent and other effects may
inhibit  collapse or even suppress it completely.

It is crucially important to establish the physical conditions within the
extended envelopes of PPCs since they will determine most of the early 
evolutionary history of the sources as they evolve towards main-sequence stars.
In particular, the initial conditions are highly important in defining the 
collapse dynamics, the likely mass of the star (and globally, the initial mass 
function), the timescales for evolution, and the likelihood of detection of
sources in different evolutionary epochs.

Different initial conditions lead to very different collapse dynamics.
Collapse from a singular isothermal sphere leads to the
``inside-out'' collapse model with a constant mass accretion rate 
\citep{Shu (1977)}. 
If collapse begins before the density distribution relaxes fully to a
singular isothermal sphere, there will be an 
initial period of much higher mass accretion rate, followed by a declining
rate of accretion
\citep{Foster and Chevalier (1993), Henriksen
et al. (1997)}. Such an epoch has been identified with the Class 0 objects
\citep{Andre et al. (2000)}. 
Collapse from a logatropic sphere, which produces mass accretion rates
that {\it increase} with time, has also been suggested \citep{McLaughlin and
Pudritz (1997)}.
To address these issues, it is necessary to understand the density and
temperature distributions within the cores and to correlate them to the
kinematics. In this paper we address the first of these issues.

In the original studies of PPCs \citep{Ward-Thompson et al. (1994), 
Ward-Thompson et al. (1999)}, the radial distribution of the continuum emission 
was found to be relatively flat toward the center of the PPCs; this 
{\it intensity} distribution was interpreted purely in terms of a {\it density} 
distribution that is relatively flat within the inner core and falls off as an 
approximate power law at larger separations. 
These authors, and others, have invoked such 
density distributions as proof of the significance of non-thermal (magnetic
and/or turbulent) pressure support in these objects. However, 
\citet{Ward-Thompson et al. (1999)} pointed out that the source statistics and, 
where available, the observed magnetic field strengths suggest a quantitative 
disagreement with the evolutionary timescales in the published ambipolar 
diffusion models for the quasi-static contraction of magnetized cloud cores.
This conclusion has been questioned by \citet{Ciolek and Basu (2001)}.

These conclusions are important in constraining the collapse process, but 
previous interpretations of the data assumed that 
the dust temperature is uniform throughout the cores. 
We will argue that the interpretations of the 
observational data are profoundly affected by this assumption.
Early calculations for dust clouds heated by the external radiation
field found that the temperature declined to very low values in
the interior \citep{Leung (1975)}; more recently \citet{Zucconi et al.
(2001)} have applied a semi-analytic technique and come to similar 
conclusions.

In this paper, we constrain the density and temperature profiles of a small
sample of PPCs, utilizing high quality observational data and a dust continuum 
radiative transfer code \citep{Egan et al. (1988)}.
We consider a range of physical models for the density distribution,
calculating the dust temperature distribution in each case
in a self-consistent way. The model clouds are then ``observed" 
to mimic the observational technique (including the effects of the 
telescope beam power pattern and chopping) in order to compare to actual 
observations. Although several sources of observational data are considered, 
the primary observational comparison is to our recent 450$\mu$m and 850$\mu$m 
data obtained with the SCUBA on the James Clerk Maxwell Telescope (JCMT) 
\citep{Shirley et al. (2000)}.

\section{Physical Models}

We consider the simplest physical situation of thermally 
supported, gravitationally bound, spherically symmetric clouds. 
While all three cores modeled here have some degree of turbulence, it is
subsonic in L1512 and L1544 or barely sonic in L1689B; 
the effects of turbulence should not be dramatic in these very quiescent cores.
The solution for the thermally-supported, isothermal cloud is the 
Bonnor-Ebert sphere, 
which can be generalized to account for a temperature gradient (see below).
In addition to the thermally supported Bonnor-Ebert spheres, we 
have also considered a singular sphere, for which  $\rho\propto r^{-2}$, 
normalized to match a Bonnor-Ebert model at the outer radius.
Magnetically supported cores have also been suggested
\citep{Ciolek and Mouschovias (1994), Ciolek and Basu (2000)}.
The latter authors have
presented density profiles that are specifically appropriate to L1544.
Although a comparison with these models would be of great interest, 
the inclusion of magnetic pressures necessarily implies significant deviations
from spherical symmetry -- in fact, highly flattened
disk-like structures are predicted.
After some attempts to capture the essence of these models in the
one dimensional models we consider here, it became clear that a fully
two dimensional treatment is required. We defer discussion of these
magnetic models to future work but
discuss the possible effects of asphericity on our modeling 
in \S \ref{caveats}. 

We have considered two types of hydrostatic pressure-balanced cloud:
\begin{enumerate}
\item ``Classical'' isothermal Bonnor-Ebert spheres, in which $(dT/dr)=0$; and
\item Modified Bonner-Ebert spheres with a gradient 
in the kinetic temperature.
\end{enumerate}

If the gravitationally bound clouds are non-magnetic, are not 
subject to any large-scale systemic velocity flows (rotation,
infall, or outflow), and are purely supported by thermal pressure, then the 
equations of hydrostatic equilibrium are as follows:
\begin{equation}
 \frac{dP(r)}{dr}= -\frac{GM(r)\rho(r)}{r^2} ,
\end{equation}
\begin{equation}
\frac{dM(r)}{dr}= 4\pi r^2\rho(r) ,
\end{equation}
where $r$ is the radius, $P(r)$ and $\rho(r)$ are the pressure and density
respectively at $r$, and $M(r)$ is the total mass enclosed within $r$.
If the clouds are isothermal, the equation of state is
\begin{equation}
 P=a^2\rho ,
\end{equation}
where $a$ is the isothermal sound speed ($a^2=kT/\mu m_H$, $k$ is the Boltzmann
constant, $\mu$ is the average molecular mass in the gas, and $m_H$ is the 
mass of the hydrogen atom).

With an appropriate change of variables to a dimensionless form, 
$D=\rho/\rho_c$ and $\xi=(r/a)\sqrt{4\pi G\rho_c}$, where $\rho_c$ is the 
central density, these equations yield a single solution, with the maximum 
value of $\xi$ ($\xi_{max.}$) as the only free parameter (cf. Foster \& 
Chevalier, 1993), which is determined by pressure balance with the surrounding 
medium. If $\xi_{max.}>6.451$, which corresponds to a center-to-edge density 
contrast of $\rho_0/\rho_c>14.3$, then the sphere is unstable to perturbations 
and subsequent gravitational collapse. The marginal case is the ``critical 
Bonnor-Ebert sphere'' (Ebert 1955, Bonnor 1956).
The solution corresponds to a family of spheres of differing 
central density [cf. Fig. 1 of \citet{Shu (1977)}]. The density structure
of the outer envelopes is closely approximated by $\rho\propto r^{-2}$, 
but the density approaches a constant value at small $r$.
For a given temperature, higher values of $\rho_c$ yield models with smaller
cores.
The most extreme form of the unstable Bonnor-Ebert sphere has an infinite 
central density (corresponding to $\xi_{max.}=\infty$). This is the singular 
isothermal sphere, in which $\rho\propto r^{-2}$ throughout. 

The modified Bonnor-Ebert spheres include the effects of radial 
temperature variations. In these cases,
\begin{equation}
 \frac{dP}{dr}=k\left[ n\frac{dT}{dr} +T\frac{dn}{dr}\right] ,
\end{equation}
and, using $\rho =\mu m_H n$, where $n$ is the number density, equation 2 becomes
\begin{equation}
 \frac{dM(r)}{dr}=(4\pi\mu m_H)nr^2 
\end{equation}
and, from hydrostatic equilibrium,
\begin{equation}
 \frac{dn}{dr} = -\left(\frac{\mu m_H G}{k}\right)\frac{M(r)n}{Tr^2} 
-\left(\frac{n}{T}\right)\frac{dT}{dr}  .
\end{equation}

In all of our models the mean molecular mass is $\mu=2.3$, consistent with
$n = n(\hh) + n(\rm{He})$. The inner radius (\rinner) is not constrained by 
observations and is simply set small enough (25 AU) that it does not affect the 
results, as shown by tests that used $\rinner = 50$ AU.
The free parameters in the isothermal Bonnor-Ebert models are thus the central 
density (\nc), the outer radius (\router) and the gas kinetic temperature (\tk).
Most models have an outer radius of 3\ee4 AU, but some have an \router\ of 
1.5\ee4 or 6\ee4 AU.

For most of our  models (see Table 1) we have used the isothermal Bonnor-Ebert 
configurations with $\tk = 10$ K. The 
density distribution from a sequence of these models (with different central
densities, \nc) is shown in Figure 1a. 
A few isothermal models were computed for different values of \tk\ to 
examine the sensitivity of the density structure to the 
temperature (Fig. 1b).  The main effect of
changing the assumed \tk\ is to change the size of the central core and
the mass (Table 1).

For all cases with $\nc \geq 3\ee4$ \cmv,
the density contrast between the edge and the center of the cores 
exceeds 14.3, implying that they are unstable to collapse. 
For the isothermal Bonnor-Ebert spheres of fixed temperature and outer 
radius, the enclosed mass is only slightly sensitive to the central density 
($n_c$). Since the photometric fluxes are largely determined by the masses 
within the beams, observations of the spectral energy distribution (SED) 
alone cannot place very strong constraints on the central densities of 
Bonnor-Ebert spheres. However, by modeling the radial intensity distribution it 
is possible to discriminate between models with different central densities, 
which have different density {\it distributions}.

In order to generate the non-isothermal Bonnor-Ebert spheres it is
necessary to pre-define the temperature profile.
In practice, this is achieved by using an appropriately defined isothermal
Bonnor-Ebert model as the input for the radiative transfer code to 
generate the dust temperature profile (see \S \ref{isrf}). 
This \Tdr\ is then used to re-calculate the 
density profiles in the non-isothermal model. Obviously, the corrected density 
profile could then be used to generate a second iteration for the temperature 
profile etc., but the magnitude of the corrections in the second and higher 
order iterations is sufficiently small for them to be neglected.
The correction for non-isothermality results in a smaller core and an
initially steeper density profile (Fig. 1c), 
caused by the second term in equation 6, with $dT/dr >0$. 
Beyond radii of about 5000 AU, the first term in equation
6 dominates, and the smaller enclosed mass (caused by the smaller core) leads
to a shallower density profile. The {\it total} mass is larger because the
mass is dominated by the outer regions.
The shallower density profile in the outer envelope decreases
the center to edge density contrast (e.g. in the case of 
$n_c=10^6$ cm$^{-3}$, $\router = 3\ee4$ AU, by a factor of $\sim$1.8).

Note that the radiative transfer code computes the radial profile of the {\it 
dust} temperature, \Tdr, whereas the equilibrium configuration is determined 
by the {\it gas} kinetic temperature, \Tkr. While efficient gas-dust coupling 
forces \Tk\ to equal \Td\  at high densities, at densities below about
1\ee4  to 3\ee4 \cmv, $\Tk\neq\Td$ \citep{Takahashi et al. (1983), Doty and
Neufeld (1997)}. Although these models are a theoretical improvement on
the isothermal Bonnor-Ebert sphere, a full calculation of the gas
energetics, including dust coupling, would be needed to make fully 
self-consistent models.

All the models of the density profiles are tabulated in Table \ref{physmods}. 
All densities are specified as the number density of particles.
The notation tBE implies a non-isothermal Bonnor-Ebert model in which 
the density structure was computed assuming that $\Tkr = \Tdr$, where $\Tdr$
was computed from the initial isothermal BE model with the same \nc\ (model
number denoted by the superscript). 
We again emphasize the distinction between \Tk\ and \Td\ by using the
term isothermal only to refer to the assumption about \Tkr\ used to 
compute the density distribution. The {\it dust} temperature in these
isothermal BE models is {\it not} constant (\S \ref{isrf}).

\section{Interstellar Radiation Field} \label{isrf}

The radiative transport code of \citet{Egan et al. (1988)}, modified
to use an arbitrary density distribution, computed \Tdr\ for each
physical model.
We first explored the dependence of the temperature profiles on the assumed 
form of the interstellar radiation field (\isrf). Figure 2 shows a comparison 
of two estimates of the \isrf. Previous calculations with the radiative
transfer code (e.g., \citet{Leung (1975)}; \citet{Zhou et al. (1990)})
have used an \isrf\ similar to that of \citet{Mathis et al. 
(1983)}, supplemented by a blackbody for the cosmic background radiation 
(hereafter ``MMP"). 
More recent analyses, using the COBE data, indicate a stronger \isrf\ in 
the infrared \citep{Black 1994}. Significant departures from the ``MMP" ISRF
occur between 5 and 400 \micron. 
We have modified the \citet{Black 1994} \isrf\ at the 
shorter \uv\ wavelengths ($\lambda < 0.36$ \micron)
using equations given by \citet{van Dishoeck 1988}, which reproduce the \isrf\ 
of \citet{Draine 1978}. The differences between the ``MMP" and the
Black-Draine models of the \isrf\
are quite substantial, reaching a factor of 1.8 in parts of the \uv,
and a factor of 13 in parts of the infrared.

Figure 3 compares temperature profiles using the two different 
models of the \isrf, with the same physical 
model: a Bonnor-Ebert sphere with a central density of 1\ee6 \cmv\ and uniform 
kinetic temperature, $T_K=10$ K. The differences in \Tdr\ are small
($\leq 1.2$ K), but the biggest differences are at the cold center of the
cloud, where a small change in \td\ can potentially have a major impact on
the emission. Comparison of the emitted fluxes calculated for different 
wavelengths and beams indicate modest changes (less than 15\%) for $\lambda > 
170$ \micron, but bigger changes at shorter wavelengths. 
In our modeling, we use the Black-Draine model as 
the standard representation of the ISRF.

\section{Dust Opacities}

Dust opacities in molecular clouds clearly differ from those in the
general interstellar medium. Observations of regions forming massive
stars \citep{van der Tak et al. (1999), van der Tak et al. (2000)}
have been well-matched by a set of opacities calculated for
grains that have grown by coagulation and accretion of thin ice mantles
for \eten5 years at a density of \eten6 \cmv, listed in column 5 of
\citet{Ossenkopf and Henning (1994)} (hereafter OH5). This model of the dust
opacity is therefore probably appropriate to the cold, quiescent cloud cores
that are the subject of this study. We adopt these opacities
as the standard model and explore the effect of using different opacities.
As discussed below, the excellent agreement between the observational data 
and our physical models strongly supports the validity of these opacity laws.

As a specific alternative to OH5, we consider the opacities in column 2 of 
\citet{Ossenkopf and Henning (1994)} (hereafter OH2), 
which include coagulation but lack ice mantles. These may be more 
appropriate in regions where young stars have heated the grains enough to
remove the mantles.
The model opacities are larger at 1300 \micron\ than in the OH5 grain
models by a factor of about 2.2 (Fig. 2). 
Both the OH5 and the OH2 dust models result in
considerably higher opacities at long wavelengths 
(by a factor of 6.4 for OH2 at 1300 \micron) 
than for models in which the grains have not undergone 
coagulation (e.g., column 1 of \citet{Ossenkopf and Henning (1994)}.

However, despite these differences, we find that the effect of the opacity
law on \Tdr\ is small; the 
largest difference is at the inner radius, where OH2 opacities give a \Td\ 
that is lower by 0.5 K than that obtained with the OH5 opacity model.

\section{Alternative Heating Mechanisms} \label{altheating}

Because the very low temperatures of the dust grains in the centers of 
dense cores have a major impact on the interpretation of the \submm\ 
emission (\S \ref{models} to \S \ref{specmodels}), it is important to consider
all other possible heating mechanisms in addition to the absorption of the
ambient ISRF. The primary source of energy deep
inside opaque cloud cores is provided by cosmic rays. We consider direct
deposition of energy in dust grains by cosmic rays, absorption of ultraviolet
photons produced by secondary cosmic ray excitation of \hh\ molecules,
and, finally, energy transfer from possibly warmer gas heated by the cosmic rays.
For the purpose of a comparative study, we consider a representative grain of 
radius 0.1 \micron, absorption properties given by 
OH5 dust, and a cosmic ray ionization rate $\zeta = 1\ee{-16}$ s$^{-1}$.
This value is near the maximum found in recent investigations \citep{de 
Boisanger et al. 1996, Caselli et al. 1998}. The most recent estimates indicate 
$\zeta = (2.6\pm1.8)\ee{-17}$ s$^{-1}$,
based on modeling of H$^{13}$CO$^+$ observations \citep{van der Tak
and van Dishoeck (2000)}, which is in good agreement with direct
measurements by the Voyager and Pioneer spacecraft \citep{Webber (1998)}.
Consequently, the value that we have adopted should provide a strong upper 
limit to the cosmic ray heating.
In each case, we compare the heat input to a grain to the radiative
cooling by a grain at a temperature of 5 K, somewhat 
lower than the value of \td\ reached at the center of the cores. Because
the dust in the code is in  radiative equilibrium, the cooling rate
 sets a lower limit on the heating rate due to the ISRF.
At this temperature the power emitted by our representative grain, obtained by 
integrating over the product of the Planck function and the emission cross 
section for OH5 dust, is 4.0\ee{-15} erg s$^{-1}$.

Firstly, we consider the direct cosmic ray heating of dust particles.
As discussed by \citet{Greenberg 1991}, the energy input is maximized if we
assume that all cosmic ray ionizations are caused by 1 MeV protons. Then,
our value of $\zeta$ implies a flux of protons of 4.8 cm$^{-2}$s$^{-1}$,
each of which can deposit 6.13 keV in the representative grain described
above \citep{Greenberg 1991}.
This results in an energy deposition rate per grain of 2.1\ee{-17} erg 
s$^{-1}$, 190 times smaller than the radiative heating.

Secondly, we consider the effect of ultraviolet photons created following the 
cosmic ray ionization of H$_2$. This process yields energetic electrons that 
are capable of collisionally exciting electronic states of H$_2$, which decay
producing a spectrum of ultraviolet photons \citep{Prasad and Tarafdar 1983}. 
Estimates 
of the flux of these photons range from 1.4\ee3 \citep{Prasad and Tarafdar 
1983} to 1\ee4 cm$^{-2}$s$^{-1}$  \citep{Greenberg and Li 1996}. A calculation 
of particular relevance is that of \citet{Cecchi-Pestellini and Aiello 1992}, 
who include the effects of a density gradient and internal extinction of the
photons. They consider a centrally condensed cloud with properties 
similar to those that we infer for \ppc s and find a maximum photon flux near 
the center of of $\Phi_\nu = 1\ee{4}$ photons cm$^{-2}$ s$^{-1}$, for 
$\zeta = 4\ee{-17}$ s$^{-1}$. Scaling to our adopted value of $\zeta$ gives 
$\Phi_\nu = 2.5\ee{4}$ photons cm$^{-2}$ s$^{-1}$ as a maximum value.
If all of these photons are at a wavelength of
120 nm [this probably overestimates the average photon energy, \citep{Gredel 
et al. 1989}], then the energy deposition per grain is 
\begin{equation}
 \Phi_\nu E_\nu Q_a\pi a^2 < 2.5\ee4 \times 1.7\ee{-11} \times 5.1\ee{-10} 
= 2.2\ee{-16} {\rm erg\ s^{-1}}, 
\end{equation}
which is 18 times smaller than the heating rate due to the external ISRF.

Finally, we consider heating of dust grains by collisions with warmer gas
particles. The energy deposition rate depends on the gas temperature, which is
poorly constrained. However, we can put a limit to the {\it total} heating
caused by cosmic ray heating of the gas by assuming that {\it all} of the 
energy input to the gas is transferred to the dust. This is clearly a gross 
upper limit since much of the energy will be radiated away in molecular 
emission. The volume rate of heating by cosmic rays 
\citep{Goldsmith and Langer 1978} is
\begin{equation}
 \Gamma_{CR} = 3.2\ee{-27} \Bigl({\zeta \over 1\ee{-16}} \Bigr) n(\hh) 
{\rm erg\ cm^{-3}\ s^{-1}}, 
\end{equation}
Assuming that
\begin{equation}
n_d = n(\hh)/3.9\ee{11}, 
\end{equation}
where $n(\hh)$ and $n_d$ are the number densities of H$_2$ molecules and dust
grains, respectively,
the upper limit of the heating rate per grain is 1.3\ee{-15} erg s$^{-1}$;
still a factor of 3 less than the heating by the ISRF.
In principle, we could also consider the heating of the gas by the ISRF in 
the outer parts of the core, but our main concern is whether or not the central
regions of the cores are as cold as our models predict. In any case, the 
density in the outer regions is too low for the gas and dust temperatures to be
well coupled. We conclude that heating by the ISRF dominates other possible
heat sources for dust grains, even deep inside an opaque core.

Recent calculations of gas and dust energetics, including the effects
of depletion on the coolants, support the conclusion that the gas cannot
substantially raise the dust temperature \citep{Goldsmith (2001)}.

\section{Modeling Observations} \label{models}

The primary observational data that the models must reproduce are the
radial profiles of intensity at 450 and 850 \micron, as presented by
\citet{Shirley et al. (2000)}. For some sources, intensity profiles at
1300 \micron\ obtained with the IRAM 30-m telescope also exist. 
We model the radial intensity distribution,
$\nInu \equiv I_{\nu}(b)/I_{\nu}(b_0)$, 
as a function of impact parameter ($b$), normalized to the 
innermost impact parameter, corresponding to 
one quarter of the beamwidth. This measure is thus 
sensitive to the normalized density and temperature {\it distributions}, while 
the SED is sensitive to the ISRF and a 
quantity that is proportional to the product of total mass within the beam and 
the opacity. Thus, modeling the {\it normalized} radial intensity distribution
and the SED separately provide roughly orthogonal constraints on different model
variables. Photometric data, summarized by \citet{Shirley et al. (2000)},
can constrain the ISRF, mass, and model opacities. Of course, photometric data
obtained with {\it different} beams at the same 
wavelength retain some sensitivity to the density and temperature 
distributions.

To model correctly the radial intensity distribution, one must calculate
the emission from a model cloud, with a self-consistent \Tdr, convolve
the emission with the observed beam shape, and ``chop'' the model in the
same way as the data were chopped during the observations. 
The calculation of \Tdr\ is done with the code of \citet{Egan et al. (1988)}.
Figure 4 shows that more centrally condensed 
cores (higher \nc) result in lower dust temperatures in the interior (as
expected), but somewhat higher dust temperatures at larger radii until all 
models converge to about the same value at the edge of the cloud.

A second code then uses these density and temperature profiles [$n(r)$ and 
\Tdr] to calculate the angular dependence of the emission at specific 
wavelengths, convolves with the observed beam, and uses a numerical 
simulation of the 
chopping that was used in obtaining the data to produce a predicted intensity
profile, normalized in the same way as the data (\nInum). 
The simulation of chopping in a one dimensional code cannot replicate
exactly the observations at large impact parameters. We limit our comparison 
of models to data to ranges of $b$ where the predictions are insensitive
to the exact model of chopping.
In addition, the ratio of the normalized intensities at the two wavelengths 
[$I^{norm}_{450}(b)/I^{norm}_{850}(b)$]
is computed to examine possible variations in the spectral index with radius.
Figure 4 shows how the predicted intensity distributions are affected by the
assumed form of the density distribution. Although the more condensed 
models lead to more rapid declines of the density with radius, 
beam convolution and chopping, as shown in Figure 4c and 4d, makes the
difference less striking than one might expect
[see also Fig. 1 of \citet{Shirley et al. (2000)}].

For the 1300 \micron\ data obtained with the 30-m IRAM telescope, 
the beam power pattern is available from the literature,
but the data cannot be completely modeled because they were obtained with
multiple chopper throws and restored. To bracket the range, we ran models
with no chopping and models with a 120\arcsec\ chop, based on an estimate
of where the IRAM data begin to lose sensitivity (P. Andr\'e, pers. comm.).
Clearly, some loss of signal at large impact parameters has occurred because the
unchopped models do not reproduce the declines seen in the multiply-chopped 
data, but chopping at 120\arcsec\ may overestimate the effect.

Modeling the SED allows tests of different aspects of the models:
the ISRF, the assumed dust opacities, and the mass within the beams.
In particular, \fir\ data constrain the ISRF, while \submm\ data constrain
the product of mass and \submm\ opacity.
Photometric data are quite variable in quality;
most references do not provide a beam shape and the data often represent
the emission integrated over a map. In these cases, we assume a Gaussian beam.
In addition, calibration uncertainties are hard to quantify.
Far-infrared data from \ISO\ exist \citep{Ward-Thompson et al. (1998)} and
\citep{Ward-Thompson and Andre (1999)}, but
the flux density estimates were being revised to
correct for background subtraction.
While this paper was being refereed, the new measurements became
available \citep{Ward-Thompson et al. (2001)}. The new results 
require an ISRF lower than used in our models, but the conclusions
about the best density model are unaffected. Consequently, only 
a few new models were calculated for each source to illustrate the
effects of the lower ISRF. These new results strongly support
our conclusions that the dust temperatures are very low in the centers
of these cores

The agreement between the model and the data can be quantified in terms
of the reduced chi-squared 
\begin{equation}
\chi_r^2 = (\nInu - \nInum)/\sigma)^2/N
\end{equation}
where $\sigma$ is the uncertainty in the data and $N$ is the number of data
points; only points spaced by a full beam are used in computing \chisq\
to avoid introducing correlations. Because the \chisq\ measure gives
great weight to a few discrepant points and/or points with small uncertainties,
we have also computed the mean absolute deviation, as used in robust 
estimation:
\begin{equation}
\mean{\delta} = \vert \nInu - \nInum \vert /N .
\end{equation}
The statistics were computed separately for each observation. Generally,
the computations were carried out to angular distances of about 60\arcsec, 
beyond which inability to simulate chopping exactly
make the comparison dubious for the 
SCUBA data. A similar calculation with the flux densities (\Snu) 
was used for comparing the observed
and model SEDs. In Tables \ref{l1689b.res} to \ref{l1512.res}, we report
the {\it sum} of the \chisq\ for  the 450 and 850 \micron\ data, as well as
\meandev, but only \chisq\ for the comparison to 1300 \micron\ data and
the SED.  The sum of \chisq\ for the SCUBA data is always dominated by the 850 
\micron\ data because of the smaller uncertainties, whereas the \meandev\ 
weights the 450 \micron\ and 850 \micron\ data more equally.
For the 1300 \micron\ data, two values for \chisq\ are given in 
Tables \ref{l1689b.res}--\ref{l1512.res}, the chopped value first, 
and the unchopped value second. In the figures, the chopped prediction for
1300 \micron\ is shown as a dashed line, whereas the solid line indicates
unchopped models.

The \chisq\ values for the SED in most table entries do not include the new
\fir\ data of \citet{Ward-Thompson et al. (2001)}; two models for
each source, at the bottom of each table, do include these data in
the \chisq\ calculation.

\section{Modelling Individual Sources} \label{specmodels}

\subsection{L1689B}

For L1689B, we assume a distance of 125 pc, rather than the traditional 
160 pc, based on improved determinations by \citet{de Geus et al. (1989)}, 
and \citet{Knude and Hog (1998)}, which are supported by the distance to
upper Scorpius \citep{de Zeeuw et al. (1999)}.
The SCUBA data
on this source \citep{Shirley et al. (2000)} have fairly good signal-to-noise
and the source shape as projected onto the plane of the sky is reasonably
circular, away from the central region.
The higher contour levels are somewhat elongated and may even have
weak multiple peaks. Thus, our spherical models can only
be approximations to the actual source. In addition to the SCUBA data,
we consider the IRAM 1300 \micron\ map \citep{Andre et al. (1996)}. As
stated above, because these data were taken with multiple chopper throws and 
restored, we cannot simulate the effects of chopping as we can for the SCUBA 
data; thus we weight the agreement with the 1300 \micron\ data much less
in deciding which models are best.

For most of the models of L1689B, whose results are summarized in Table 
\ref{l1689b.res}, we
adopt the simple, isothermal Bonnor-Ebert description of the
density distribution. In these models, the primary variable is the
degree of central condensation, measured by \nc.
The inner radius was fixed at 25 AU, the outer radius at 3\ee4 AU, and the 
dust opacity was OH5.
The \chisq\ and $\mean{\delta}$ values for the SCUBA data show a clear minimum 
around $\nc = 1\ee6$ \cmv; the 1300 \micron\ data favor 
$\nc = 1-3\ee6$ \cmv.
The model with $\nc = 1\ee6$ \cmv\ is a good compromise,
and we adopt that value as our standard model 
in testing the effect of other parameters.
We also explored the effects of using non-isothermal
Bonnor-Ebert models, which may be a more realistic description of
the density distribution. The differences in the predicted intensity
distributions were very small.

The effects of chopping make the data quite insensitive to the outer radius. 
A model with an inner radius of 50 AU was essentially indistinguishable
from the standard model, as was a model with $\router  =  6\ee4$ AU.
Decreasing the
outer radius to 1.5\ee4 AU has very little effect on the SCUBA 
or 1300 \micron\ fits.

We have also considered the effects of adopting different dust opacities. 
Models using OH2 dust opacities differed little from those with OH5 dust
opacities in the fits to the SCUBA and the 
1300 \micron\ data, but the agreement with the SED worsened 
substantially. Because OH2 dust has a higher opacity at long wavelengths,
these models produced too much flux. The excess flux could not be decreased 
by varying
\nc\ within the range favored by the intensity profiles because the amount of 
mass in the beam changes little with \nc. It is remarkable in fact that
OH5 dust opacity and a Bonnor-Ebert density distribution fits the SED so well. 
If we were sure that the cloud is hydrostatic and thermally pressure-balanced,
so that a Bonnor-Ebert sphere is the correct physical model, then we 
could constrain the dust opacities quite strongly.

The new \fir\ data \citep{Ward-Thompson et al. (2001)} are about half the
values previously reported, indicating lower \td, and thus a lower value
for the ISRF than assumed in our models. The last two models in Table
\ref{l1689b.res} include these new data. The first retains the original 
ISRF, while the second scales the ISRF by a factor of 0.5, clearly 
improving the \chisq\ for the SED. The latter model is shown in Fig. 5,
including the temperature profile, the
450/850 $\mu$m intensity ratio, the SED, and fits to the 450, 850, and 
1300 $\mu$m normalized intensity profiles.

Given the substantial
uncertainties, the model fits the data well. In particular, the data 
suggest (with big uncertainties) a rise in the ratio toward the edge of the 
cloud. The model correctly predicts this behavior which arises almost equally 
as a result of the increase in \Td\ at the edge of the cloud and as an effect 
of chopping.
In principle, deviations of the data from the model could be used to study 
possible changes in dust properties with radius, but better signal-to-noise in 
the 450 \micron\ data is needed before it is possible to make 
quantitative analyses. 
For the present, we note that any attempt to model the distribution of
dust properties must incorporate careful considerations of spatial variations
in the dust temperature and the effects of chopping.
We can also compare the actual values of the spectral index $\alpha$ to the
observations \citep{Shirley et al. (2000)}.
The best model produces $\alpha=2.1$ in the
40\arcsec\ beam and $\alpha=2.4$ in the 120\arcsec\ beam (see \citet{Shirley
et al. (2000)} for the definition of $\alpha$. These values
are consistent within uncertainties with 
the $\alpha$ seen in the observations ($2.0\pm0.6$).

We may ask how our conclusions compare to the previous studies of 
\citet{Andre et al. (1996)} (AWM).
Firstly, AWM assume a distance of 160 pc and an opacity at 1300 \micron\
of $\kappanu = 0.005$ cm$^2$ gm$^{-1}$ of gas, whereas the corresponding value
for the OH5 dust opacity is $\kappanu = 0.009$ cm$^2$ gm$^{-1}$ (assuming a 
gas to dust ratio of 100). In comparing results, we adjust the results of AWM 
to our assumed distance and opacity. The other main difference is that AWM 
assumed a constant $\Td = 18$ K, whereas we compute $\Tdr$ and find it to be 
lower everywhere, especially toward the center of the cores. Because the 
lower \Td\ depresses emission from the center substantially (reducing the 
temperature from $\Td = 18$ K to $\Td =7.5$ K results in a flux reduction 
by a factor of 4 at $\lambda=1300$\micron ), we find higher central 
densities (by factors 
of 3--10) and a smaller radius (by factors of 0.5--0.9) for the flattened part 
of the density distribution. 
The smallest differences are found in the model with $\nc = 3\ee5$ \cmv .
The mass inside \eten4 AU is more similar; our mass (0.95 \msun)
is 1.4 times higher than that of \citet{Andre et al. (1996)}.
Both studies agree that the density distribution outside the flat
portion is well represented by a power law ($n(r) \propto r^{-p}$)
with $p \sim 2$.

\citet{Bacmann et al. (2000)} have re-analyzed the 1300 \micron\ data,
assuming a constant \Td\ of 12.5 K and incorporating an analysis of
the absorption at $\lambda \sim 7$ \micron\ that the cloud shows in
ISOCAM data. They also find a density distribution that shows a distinction
between a ``flat'' inner region and an envelope in which the density
distribution can be described by a power law, roughly consistent with $p = 2$.
They also find that Bonnor-Ebert spheres fit their data quite well. 
However, they find that the 
radius of the flat density core is 5000--6000 AU, 3 times larger than
we find. This \rflat\ would correspond
roughly to our Bonnor-Ebert sphere with $\nc = 1\ee5$ \cmv, which
fits neither our SCUBA data nor the 1300 \micron\ data, in our analysis. 

In principle, the absorption method does not depend on \Td, thereby
providing a complementary probe. However, the complication arises in the
determination of the foreground emission, which veils the absorption by the
core. \citet{Bacmann et al. (2000)} use the 1300 \micron\ data to 
constrain this foreground emission and conclude that the foreground is
almost entirely zodiacal emission, so that the cloud must be illuminated
entirely from behind. With this assumption, they find $\tau(7 \micron) = 0.7$;
if there is more foreground veiling, the cloud could have $\tau > 1$  at
7 \micron, as our models indeed predict. 
Another variable is the opacity at 7 \micron. \citet{Bacmann et al. (2000)} 
assume an opacity that translates to $3.1$ cm$^2$ gm$^{-1}$ of gas, whereas the
OH5 dust opacity would imply $12.5$ cm$^2$ gm$^{-1}$.
\citet{Bacmann et al. (2000)} also find evidence from the density profiles 
for an outer cut-off radius to the core. Since the core is flattened, this
cut-off radius is variable, but taking the geometric mean of the two axes 
and correcting to a distance of 125 pc yields a radius of $2.6\ee4$ AU,
similar  to our outer radius, but outside the region that we effectively probe.
While it is difficult to sort out all of these issues, it seems
likely that the discrepancy between our models and those of 
\citet{Bacmann et al. (2000)} imply that either they have underestimated
the foreground veiling and hence $\tau(7 \micron)$ or we have underestimated
the dust temperatures. The new results in the \fir\ are well matched
by our low \Td, supporting our interpretation. Because the absorption
measurements provide an independent measure of the distribution, this
difference leaves a puzzle for further work.

We conclude that Bonnor-Ebert spheres and OH5 dust opacities reproduce the
observations of L1689B very well, with $\nc\ \sim  1\ee6$ \cmv.
This density is higher and \rflat\ is  smaller than
have been found in other studies.

\subsection{L1544}

L1544 is at a distance of 140 pc \citep{Elias (1978)}. 
The SCUBA data have slightly better
signal to noise than those of L1689B. While the source is clearly elongated
\citep{Shirley et al. (2000)}, we model it as a sphere. Since the radial
profiles are spherical averages, this is the most consistent comparison
for the present. Ultimately, aspherical models should be constructed, but
we address the likely effects in \S \ref{caveats}. In addition,
we model 1300 \micron\ data, subject to the same caveats as for L1689B:
in particular, we do not expect to fit the intensity profile at large impact 
parameters.
Magnetic collapse models have been developed for this source \citep{Ciolek and 
Basu (2000)}. Using their semi-analytic calculation of dust temperature,
\citet{Zucconi et al. (2001)} find good agreement between this model
and the data. Because we focus on spherically averaged data here, we
do not attempt to compare this magnetic model to our data.

The results are summarized in Table \ref{l1544.res}. First, considering
only simple Bonnor-Ebert spheres, a model with $\nc = 1\ee6$ \cmv\
has the lowest \chisq\ for the SCUBA data, as well as the 1300 \micron\
data and the SED. The model with $\nc = 3\ee6$ \cmv\ has nearly as good a
value of $\mean{\delta}$, but is slightly worse on the other measures. Density
profiles appropriate to non-isothermal Bonnor Ebert spheres fit 
somewhat worse for $\nc = 1\ee6$
\cmv, but slightly better for $\nc = 1\ee7$ \cmv\ than their isothermal
counterparts. 

Most of the models did not include any \fir\ data in the calculation
of the SED \chisq\ value.
The last two models in Table \ref{l1544.res} include the new \fir\
data in the SED \chisq\ value. As with L1689B, the fit is greatly 
improved with a lower ISRF.
The isothermal Bonnor-Ebert model with the ISRF decreased by 0.6 and 
$\nc = 1\ee6$ \cmv\ is shown in Figure 6. 
The spectral index, $\alpha$, agrees well
with the observations. 

Because models with very high \nc\ do not fit the data too badly,
we have also considered the extreme limit of a singular sphere ($\nc
= \infty$), normalized to have the same density at $r_o$ as 
the Bonnor-Ebert sphere with $\nc = 1\ee7$ \cmv. This power law model has
the density distribution of 
a singular isothermal sphere at a temperature of 10.4 K \citep{Shu (1977)}.
We considered two
models of this type: the first (iPL2) assumed a constant dust 
temperature of 10 K. This model produced intensity profiles that are too strongly
peaked in the inner core, confirming the conclusions of \citet{Ward-Thompson
et al. (1999)} that power law density distributions {\it with constant \Td\ }
cannot fit the data. 
A second model (tPL2) used the radiative transfer code to compute \Tdr\
self-consistently. Surprisingly, this model fits the SCUBA data reasonably well.
While the fit to the 1300 \micron\ data was clearly worse, our data
cannot rule out 
a simple power law density model {\it with a self-consistent \Tdr\ } 
for L1544, given the current uncertainties in the data and the modeling. 
Figure 7 compares the results for the two power law models; it shows how 
important it is to include the self-consistent \Tdr\ in analyzing intensity 
distributions.

As with L1689B, we compare our results to those obtained from the analysis of 
1300 \micron\ emission \citep{Ward-Thompson et al. (1999)} and
7 \micron\ absorption \citep{Bacmann et al. (2000)}. The assumed distance
is the same in all three studies, but (as for L1689B) the opacities differ.
\citet{Ward-Thompson et al. (1999)} find $\nc = 1\ee6$ \cmv\
[using the OH5 value of $\kappanu(1300)$], while \citet{Bacmann et al. (2000)}
find $\nc = 4\ee5$, or $1\ee5$ if OH5 opacities are correct at 7 \micron.
The radius of the flat-density-profile inner core ranges from 0 to 1600 AU in
our models, compared to 2500 AU \citep{Ward-Thompson et al. (1999)} or 
2900 AU \citep{Bacmann et al. (2000)}. The latter authors also identify an 
outer (cut-off) radius of 8900 AU.  This is close to the size where our
chopping and source asphericity makes it difficult to constrain the 
density distribution.
As with L1689B, our results are qualitatively similar to other analyses,
but differ quantitatively; the differences suggest either that
we have underestimated \Td\ or that the optical depth at 7 \micron\ has
been underestimated. 

As for L1689B, Bonnor-Ebert spheres and OH5 dust fit the data well.
Our most striking conclusion that is different from the
other studies is that our data {\it do not} rule out a singular power-law
density distribution.

\subsection{L1512}

L1512 is also at a distance of 140 pc, but the \submm\ emission is
much weaker than from L1544. While the observational constraints are
clearly much weaker than for the other sources, it is interesting to
model L1512 because it seems to have a flatter distribution of intensity
than the other sources. \citet{Shirley et al. (2000)} presented only
the 850 \micron\ data, but the 450 \micron\ data does have a definable
intensity distribution. Both the 450 \micron\ and
the 850 \micron\ data and some photometry constrain the models. 
The photometry was collected
in tables in \citet{Shirley et al. (2000)}, except for a datum at 1300
\micron\ from \citet{Ward-Thompson et al. (1999)}.

The results for L1512 are given in Table \ref{l1512.res}.
For this source we have only considered simple Bonnor-Ebert
density distributions --- the poor data quality precludes the consideration of
more sophisticated models.
The best fit is for $\nc = 1\ee5$ \cmv, with $\nc = 3\ee4$ \cmv\ also 
acceptable.  L1512 is clearly less centrally
condensed than the other cores. 

The last two models in Table \ref{l1512.res} include the new \fir\
data in the SED \chisq\ value. The fit is clearly very bad for the
full ISRF; the fit is much improved for an ISRF decreased by 0.3,
but it is still worse than for L1689B and L1544. The last model
is shown in Fig. 8.

The problem with the SED is that
the model fluxes are all higher than the observed values. However,
good fits are obtained for the intensity profiles and the spectral index is 
consistent with the observations.
Models with lower \nc\ fit the lower fluxes better, but do not fit the 
intensity distribution. If the dust opacities are the same as in the other 
sources, fitting to the SED requires the core to be less massive.
Bonnor-Ebert spheres calculated with $\tk = 5$ K contain less mass, and
a model with $\nc = 3\ee4$ \cmv\ fits both the SCUBA data and the SED
best (Table \ref{l1512.res}). However, such models are hard to justify on
physical grounds; the radiative transport calculation
produces $\Tdr > 12$ K everywhere (a model with $\tk = 15$ K fits the
SED very poorly). While the
density is too low to require $\Tk = \Td$, one would have to invoke unusually
low gas heating to produce $\Tk = 5 $ K. Another way to decrease the 
emission from the model cloud is to use different dust opacities; values
between those of OH5 and uncoagulated grains [e.g., column 1 of 
\citet{Ossenkopf and Henning (1994)}] would give about the right amount
of emission. Such grains are plausible; if L1512 is a younger source,
it will have spent less time at densities that are sufficiently high for grain
coagulation to occur.

L1512 was detected at 1300 \micron\ by \citet{Ward-Thompson et al. (1999)}
but was too weak to map. It was also included in the absorption study
by \citet{Bacmann et al. (2000)}, but the absorption was too weak to analyze.
These results are consistent with our finding that the mass and central density
are less than in L1689B and L1544.

The radial profiles of L1512 can be matched if the source is 
a Bonnor-Ebert sphere with a central density smaller than those of L1689B
and L1544. 

\section{Caveats and Future Work} \label{caveats}

The fact that the masses of Bonnor-Ebert spheres, together with the OH5
dust opacities \citep{Ossenkopf and Henning (1994)} match the
\submm\ fluxes so well suggests that both the physical and
dust models may be reasonable. However, the opacity and mass effectively
enter as a product; consequently, other mass and opacity combinations cannot
be ruled out.

Another caveat is that our models are one-dimensional.
Some of the cores (L1544 most notably) are clearly aspherical and (at least)
two-dimensional radiative transfer models should be used to interpret the data.
We can get a crude estimate of the effect of asphericity on the 
temperature distribution. Comparing \Tdr\ for models with the same
\nc\, but a factor of 2 smaller \router\ simulates the effects on heating
of having one dimension smaller than the other. In fact, the differences
are very small except near the surface of the cloud 
($\leq 0.2$ K, or 3\%, for $r \leq 1\ee4$ AU). 
Semi-analytic calculations of \Td\ in a two-dimensional geometry 
\citep{Zucconi et al. (2001)} find modest differences in the \Tdr\
along long and short axes.

Differences in the assumed density law will produce different \Tdr\
in general, as can be seen from Fig. 4. However, if one plots the
\Td\ versus column density, the behavior is much more universal,
as would be expected. This is shown in Fig. 9 for a variety of density
profiles. The curves of \Tdr\ track each other until they approach their
maximum value, where geometrical effects near the center of the cloud
cause them to deviate from models with higher total column density.
This figure may be useful for those modeling different
density distributions who do not have access to a radiative transport
code. For example, the $\Td \sim 10$ K computed from gray-body fits
to the SED \citep{Ward-Thompson et al. (2001)} for L1689B and L1544
would correspond to the model \Td\ at about 1/3 the total column density
into the cloud; in contrast, the radius at which $\Td = 10$ K is about
$4\ee3$ AU (Figs. 5 and 6), about 0.13 of the outer radius, 
enclosing 0.15 of the total mass.
Clearly, gray-body fits to a constant \Td\ weight radius, column density,
and mass differently, depending on the detailed density and temperature
distributions.

Local variations in the ISRF are another source of uncertainty. If local
sources increase the strength of the ISRF, \Td\ will increase. However, in
practice, large changes in the ISRF strength are needed for significant
variations to occur. Approximately, $\Td \propto L^{1/(\beta +4)}$, 
where $L$ is the ISRF luminosity and $\beta$ is the
power-law index for the opacity ($\kappanu \propto \nu^\beta$) 
\citep{Doty and Leung (1994)}.
For typical values of $\beta$, the exponent  is only 0.2 to 0.25, so
that doubling the ISRF strength will increase the dust temperature by just
15-20\%.
Full radiative transfer models run with higher ISRF strengths confirm this 
simple analysis.
Figure 3 shows the \Tdr\ calculated for the same physical model, but
with the Black-Draine ISRF multiplied by a factor of 2 (except for the
cosmic background radiation). 
The temperatures are higher at all radii by
a factor around 1.14, as would be predicted by the simple relations above.

In fact, the new \fir\ data indicate a lower ISRF than we have assumed,
even for L1689B, where the ISRF is generally considered to be enhanced.
Thus lower \Td\ are more likely than higher. Since \citet{Ward-Thompson
et al. (2001)} describe their results in terms of {\it enhanced} ISRF, a word
of explanation is in order. They use the ISRF of \citet{Mathis et al. (1983)},
which is considerably lower than the Black-Draine field that we use (cf.
Fig. 2a) in the \mir\ to \fir\ region where they constrain the field.
It seems that, for these cores at least, the correct value lies between
the two standard choices for the ISRF. The \fir\ data strongly
constrain the ISRF because they are exponentially sensitive to \Td\
in this regime; future measurements and modeling will allow much
better knowledge of the local ISRF.

\section{Conclusions}

We have modeled the emission from dust in \ppc s, including a 
self-consistent calculation of the temperature distribution (\Tdr)
for an input density distribution. We have also simulated the observations
by convolving models with the observed beam and applying chopping 
to the models.
Using the calculated \Tdr\ has a substantial impact on the conclusions. 
Compared to earlier studies
that assumed a constant \Tdr, our models indicate smaller regions of
relatively constant density (by factors of 0.5 to 0.9) and higher central
densities (by factors of 3 to 10). Indeed, for L1544, a singular, power-law
density distribution cannot be ruled out. 

For the three sources we have modeled, there seems to be a sequence of
increasing central condensation, from L1512 to L1689B to L1544. 
While many more starless cores need to
be modeled, it is possible that a new sequence of cores may be
identified, in which increasing central condensation is the primary
variable. The denser two of these cores, L1689B and L1544, have 
spectroscopic evidence of contraction motions \citep{Tafalla et al. (1998),
Gregersen and Evans (2000)}, while L1512 does not \citep{Gregersen and
Evans (2000)}, consistent with this suggested sequence.

It is interesting that Bonnor-Ebert spheres fit the data well, even
though unstable spheres are needed in all cases. Magnetic fields
may allow these nominally unstable objects to persist, but in the absence of a
suitable 2D/3D radiative transfer model we are not able to say whether or not
the existing ambipolar diffusion controlled collapse models are consistent
with the observations.

\citet{Johnstone
et al. (2000)} have also used Bonnor-Ebert spheres to fit their SCUBA data 
in the $\rho$ Ophiuchus molecular cloud. They find smaller outer radii
and smaller degrees of central condensation, but they have used 
models with constant $\Td = \Tk$. 
The smaller outer radii may reflect the crowded conditions and
higher ambient pressures in the $\rho$ Ophiuchus region, compared to the 
``elbow room" available to the isolated cores in this study. More detailed
analysis of the clustered regions, including calculations of \Tdr, and
more extensive studies of isolated regions will together delineate the
differences in initial conditions for clustered and isolated star formation.


\acknowledgements

We are grateful to the referee, C. Wilson, for helpful suggestions and to 
G. Ciolek and S. Basu for discussion of magnetic models.
We are very grateful to 
P. Andr\'e and D. Ward-Thompson, who provided data and helpful discussion.
We thank E. van Dishoeck for advice on the interstellar radiation field and
M. Greenberg for helpful discussion on alternative grain heating mechanisms.
T. Greathouse helped with some of the computer codes.
The JCMT is operated by the Joint Astronomy Centre on behalf of the Particle 
Physics and Astronomy Research Council of the United Kingdom, The Netherlands 
Organization for Scientific Research and the National Research Council of 
Canada.
We thank the State of Texas and NASA (Grant NAG5-7203) for support.
NJE thanks the Fulbright Program and PPARC for support while at University
College London and the Netherlands Organization for Scientific Research (NWO)
bezoekersbeurs grant and NOVA for support at Leiden Observatory.

\newpage

\begin{figure}[hbt!]
\plotfiddle{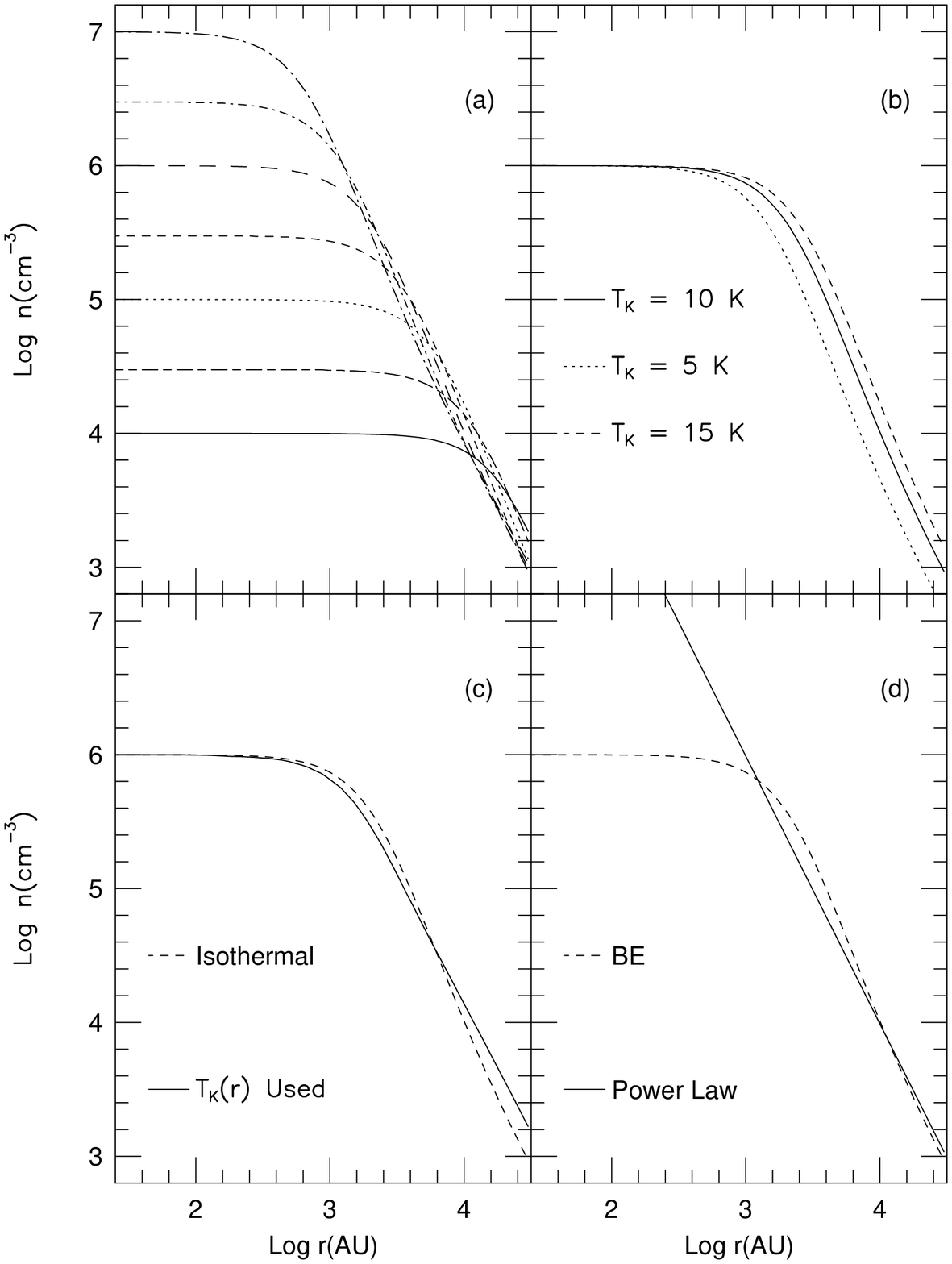}{6.5in}{0}{65}{65}{-200}{-20}
\caption{Plots of log density versus log radius in AU for the input
physical models. In (a), Bonnor-Ebert spheres with constant kinetic 
temperature, $\Tk = 10$ K and central densities from
$\nc = 1\ee4$ to $\nc = 1\ee7$ \cmv\ are shown. In (b), Bonnor-Ebert spheres
with $\nc = 1\ee6$ \cmv\ are shown for different, constant kinetic 
temperatures. In (c), a Bonnor-Ebert sphere with a variation in \Tk,
based on iteration with the radiative transport code, and assuming
$\Tkr = \Tdr$, is compared to an isothermal ($\Tk = 10$ K) Bonnor-Ebert 
sphere, with the same \nc. In (d), a power law (PL2) corresponding to 
a singular isothermal sphere with $\tk = 10.4$ is compared to a Bonnor-Ebert
sphere with $\nc = 1\ee6$ \cmv.
\label{figphysmod}}
\end{figure}

\begin{figure}[hbt!]
\plotfiddle{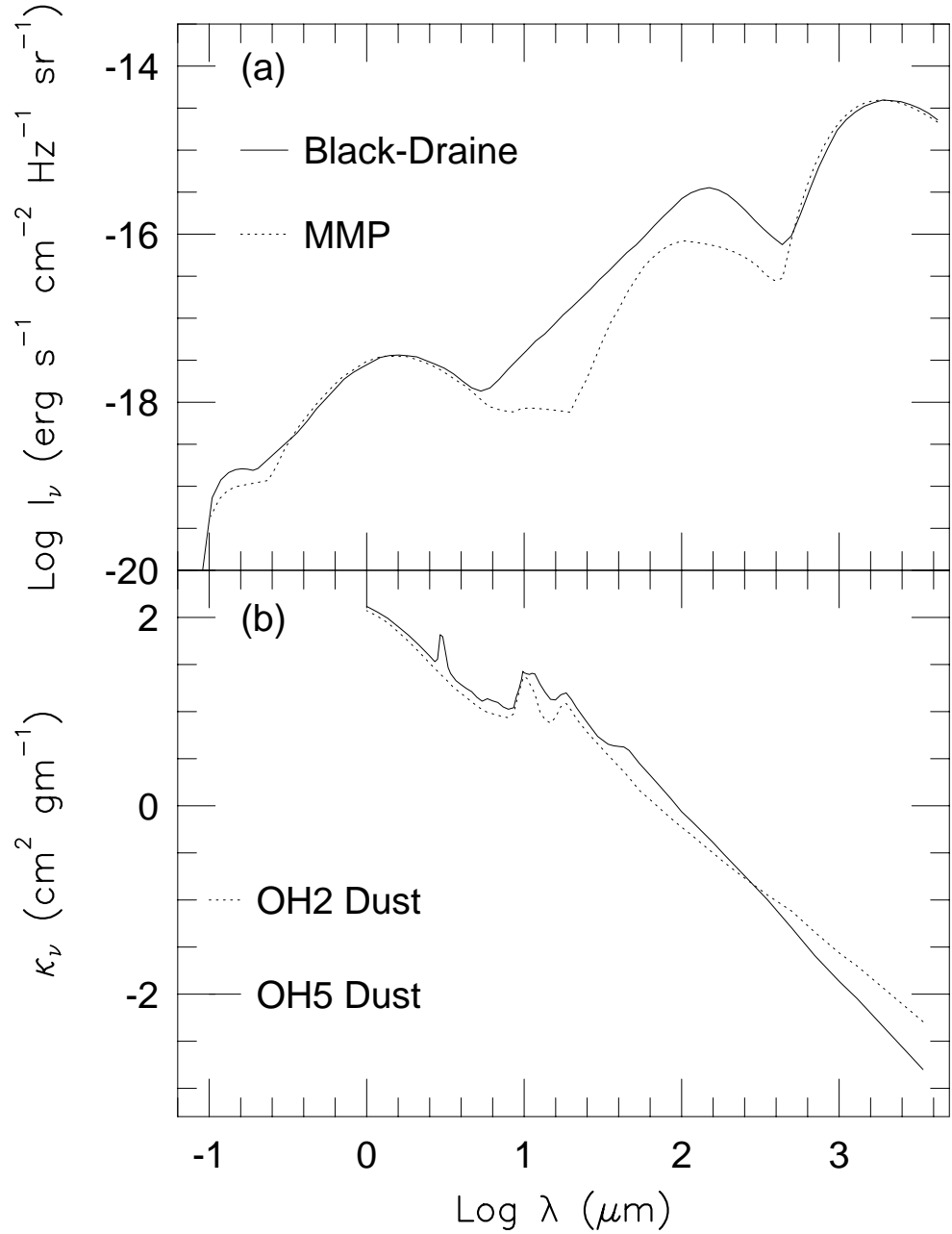}{7.0in}{0}{70}{70}{-200}{-20}
\caption{ In (a),
 the interstellar radiation field used in previous
versions of the code (``MMP") is compared with that used here (Black-Draine).
``MMP" is close to the ISRF used by Mathis, Mezger, \& Panagia (1983),
supplemented by a blackbody for the cosmic background radiation. The
ISRF labeled Black-Draine uses the curve in Black (1994) for 
$\lambda \ge 0.36$ \micron, and Draine (1978) for $\lambda < 0.36$ \micron.
In (b), the opacity per gram of gas (\kappanu) for 
OH5 and OH2 dust, based on \citet{Ossenkopf and Henning (1994)} and
a gas to dust mass ratio of 100, is plotted versus wavelength.
\label{figisrf}}
\end{figure}

\begin{figure}[hbt!]
\plotfiddle{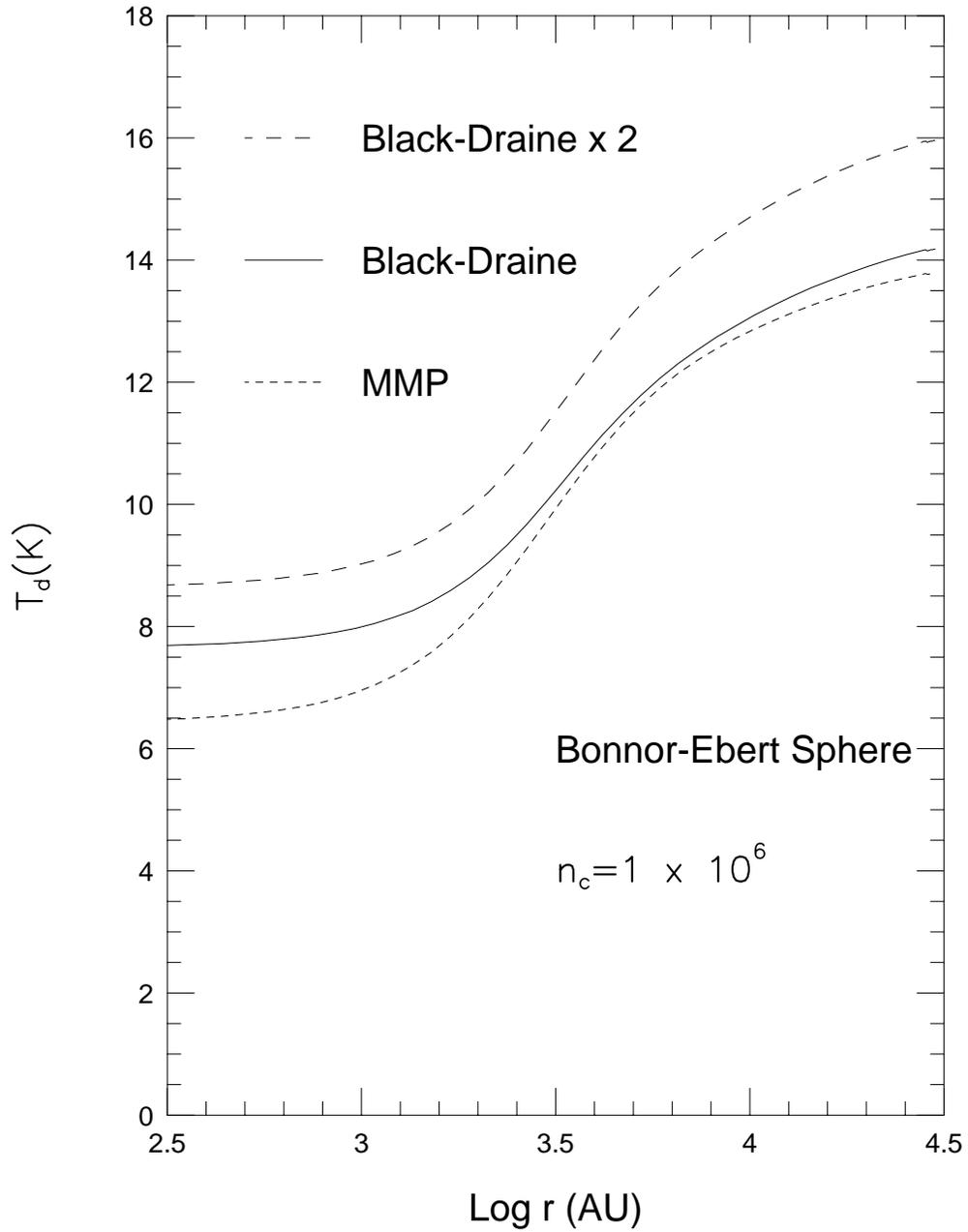}{7.0in}{0}{70}{70}{-200}{-20}
\caption{Shows the effects on the dust temperature distribution
of changing the ISRF from ``MMP" to Black-Draine, and of using twice the
average ISRF. A Bonnor-Ebert sphere with $\nc = 1\ee6$ \cmv\  and
$\Tk = 10$ K was used for all three calculations. \label{figtdisrf}}
\end{figure}

\begin{figure}[hbt!]
\plotfiddle{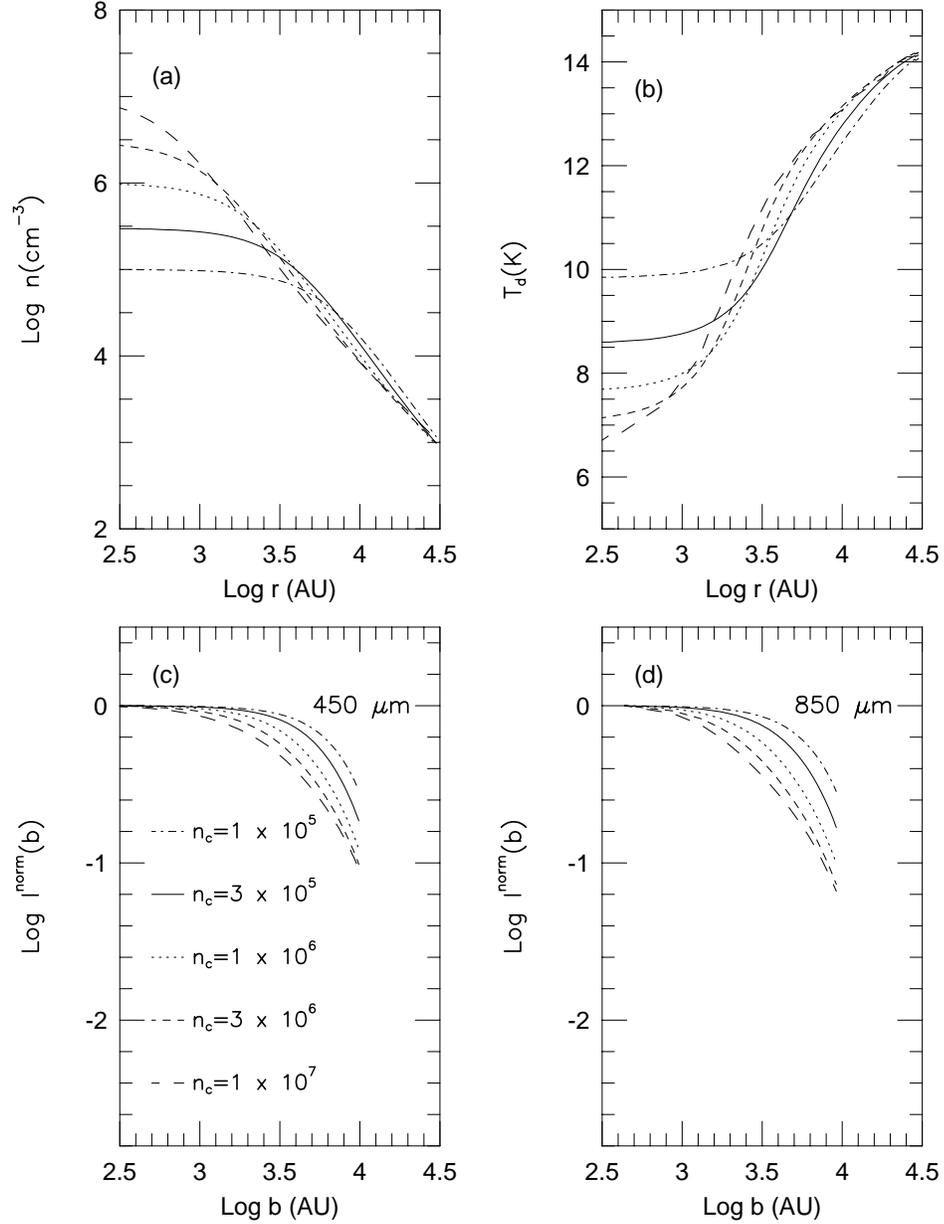}{7.0in}{0}{70}{70}{-200}{-20}
\caption{The density distributions for a series of Bonner-Ebert spheres
are shown in (a). The resulting temperature distributions are shown
in (b), with the same line coding; the densest models have the lowest
central temperatures. In (c), the resulting intensity distributions
at 450 \micron\ are shown. In (d), the results for 850 \micron\ are
shown. The same line coding is used for all panels, with the densest
models showing the fastest fall-off with radius. The model emission
in panels c and d has been convolved 
with the observed beam and chopping by 120\arcsec\ 
has been simulated, causing some of the drop around impact parameter, 
$b \sim 10,000$ AU, for a distance of 125 pc.
\label{figmod}}
\end{figure}

\begin{figure}[hbt!]
\plotfiddle{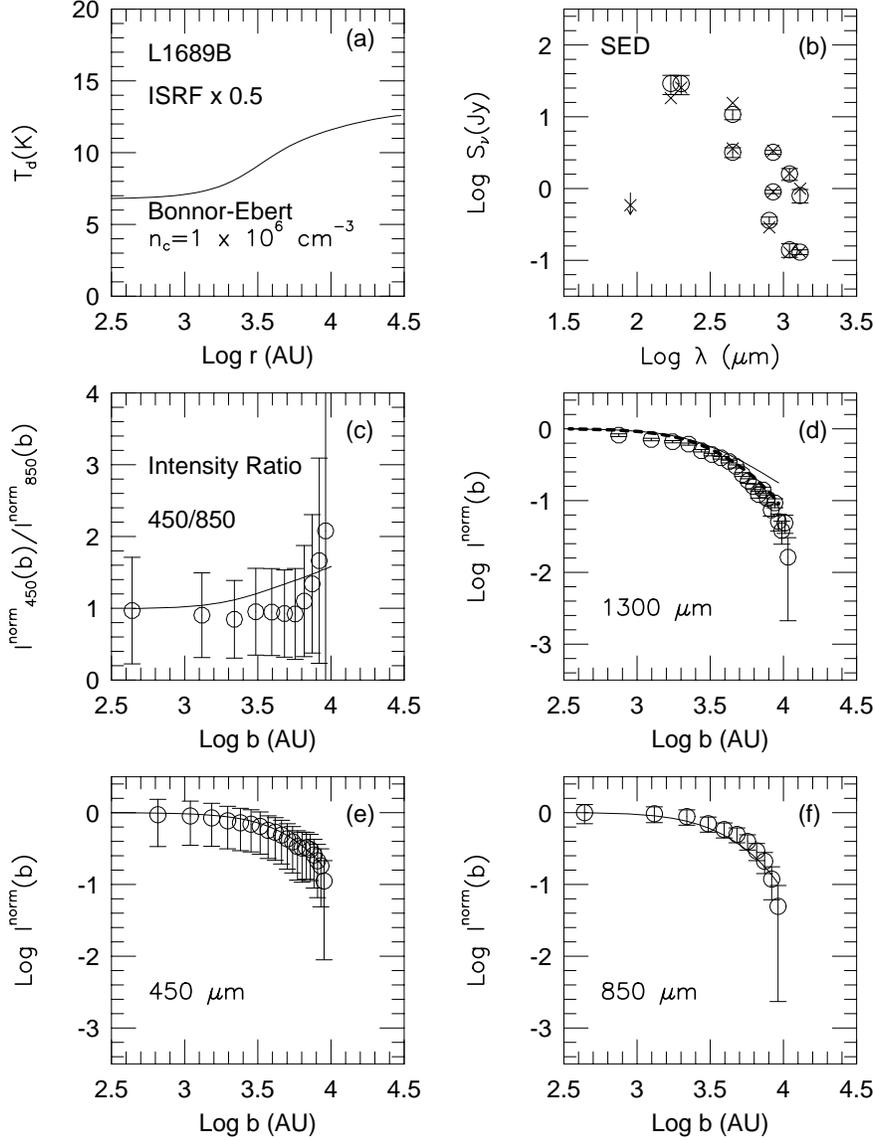}{6.5in}{0}{65}{65}{-200}{-20}
\caption{A model for L1689B that fits the data well, a Bonnor-Ebert sphere
with $\nc = 1\ee6$ \cmv\ and a Black-Draine ISRF multiplied by 0.5. 
The temperature distribution is shown in (a). 
The observed SED is shown in (b) as circles with errorbars and an upper limit
at 90 \micron; multiple values at the same wavelength are data with different
beams. The crosses are the model predictions for the same beams.
The bottom two panels, (e) and (f), show for 450 and 850 \micron,
the observed, normalized intensity profile (circles and error bars)
and the model (solid line), with simulated chopping. Panel (d) shows
the data at 1300 \micron\ of Andr\'e, Ward-Thompson, \& Motte (1996) and
the model without simulated chopping (solid line) and with a simulation
of chopping by 120\arcsec\ (heavy dashed line). Panel (c) shows the ratio of 
the 450 and 850 \micron\ normalized intensities with the same conventions. Note
that normalization constrains the value to unity at the innermost point.
At 125 pc, $b = 10^4$ AU corresponds to 80\arcsec.
 \label{figl1689b}}
\end{figure}

\begin{figure}[hbt!]
\plotfiddle{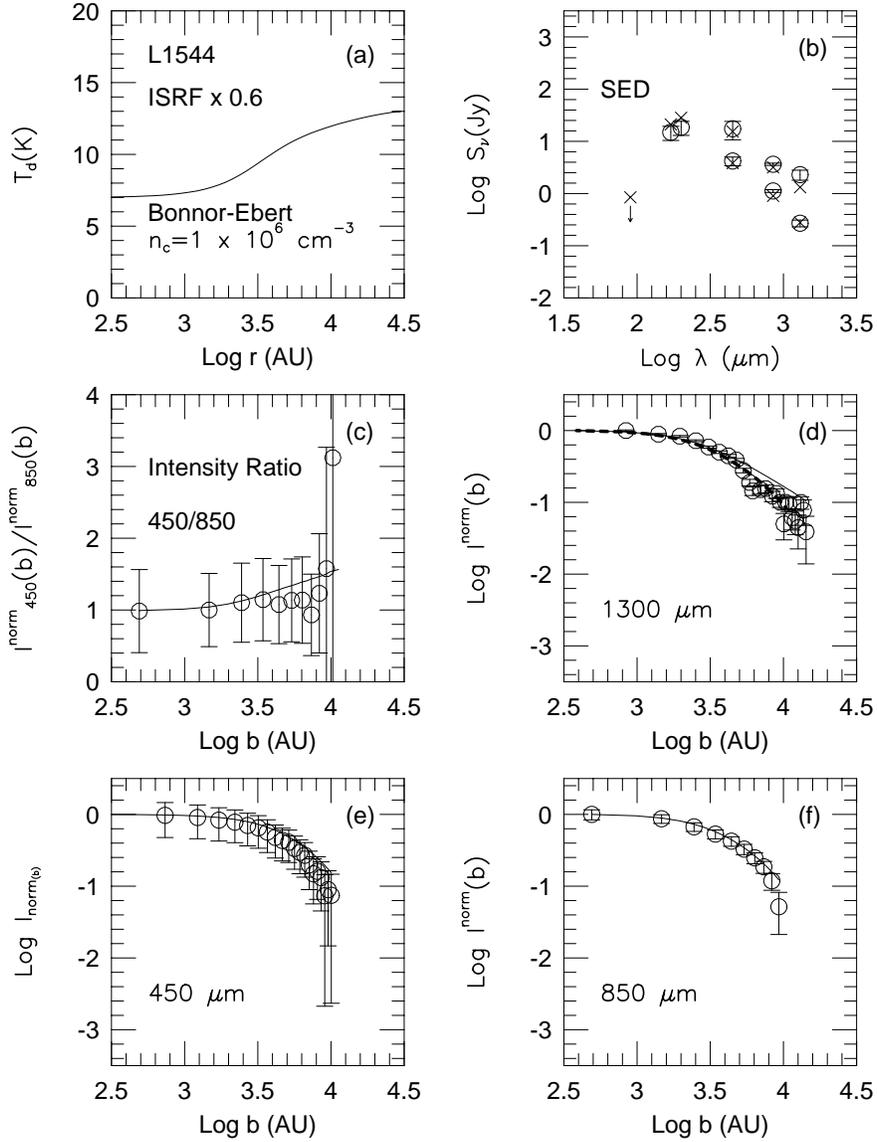}{6.5in}{0}{65}{65}{-200}{-20}
\caption{The same as Fig. \ref{figl1689b}, but for L1544. The model is
a Bonnor-Ebert sphere with $\nc = 1\ee6$ \cmv\ and a Black-Draine 
ISRF multiplied by 0.6. The 1300 \micron\ 
data is from Ward-Thompson, Motte, \& Andr\'e (1999).
At 140 pc, $b = 10^4$ AU corresponds to 71\arcsec.
 \label{figl1544}}
\end{figure}

\begin{figure}[hbt!]
\plotfiddle{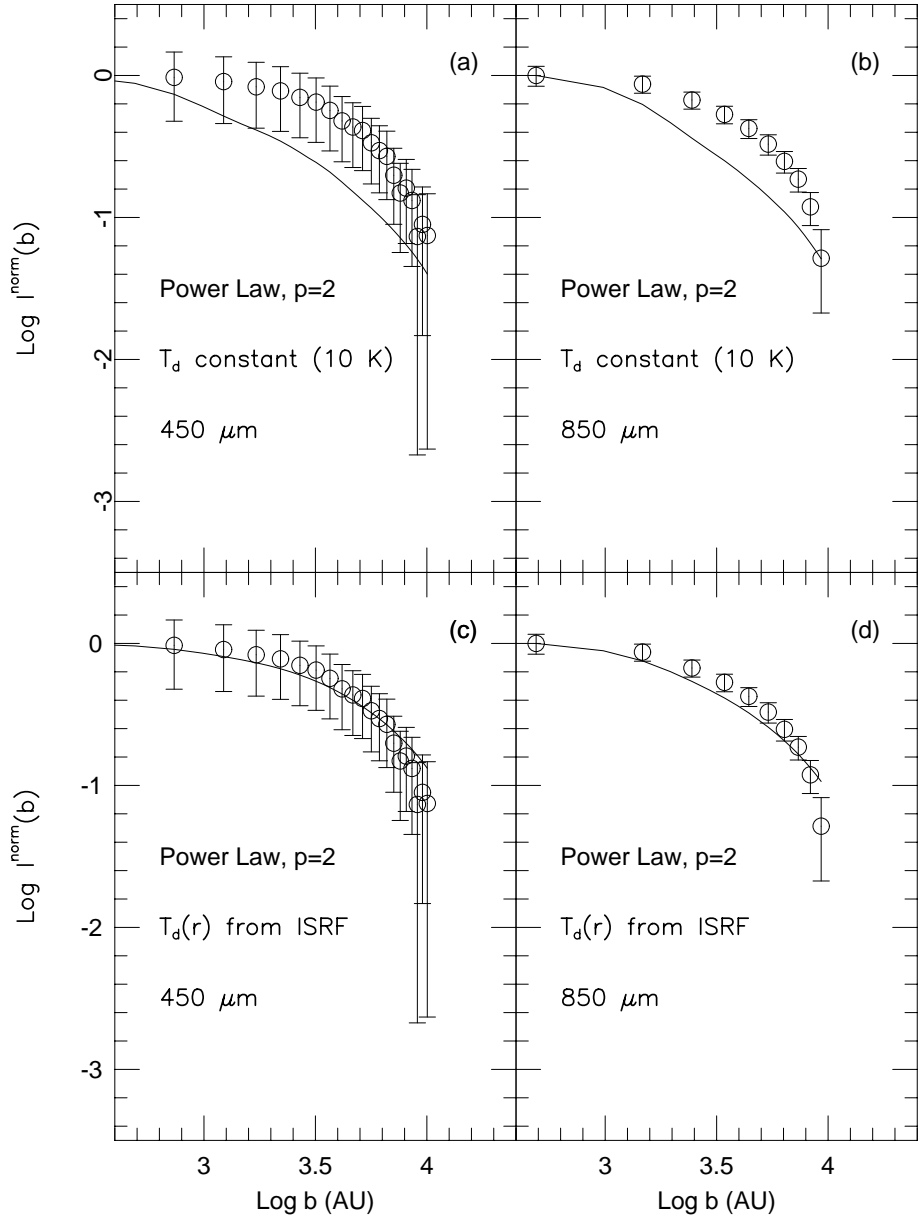}{7.0in}{0}{70}{70}{-200}{-20}
\caption{The normalized intensity distributions at 450 (a and c) and
850 (b and d) \micron\ for a power law density distribution, $n(r) = n(\router)
(r/\router)^{-p}$, with $n(\router) = 1.02\ee3$, $p = 2$, and 
$\router = 3\ee4$ AU. The observations are of L1544.
The top panels (a and b) show the model predictions with a constant
$\Tdr = 10$ K, while the bottom panels (c and d) show predictions for
\Tdr\ calculated self-consistently with the radiative transport code,
using the full Black-Draine ISRF.
 \label{teffect}}
\end{figure}

\begin{figure}[hbt!]
\plotfiddle{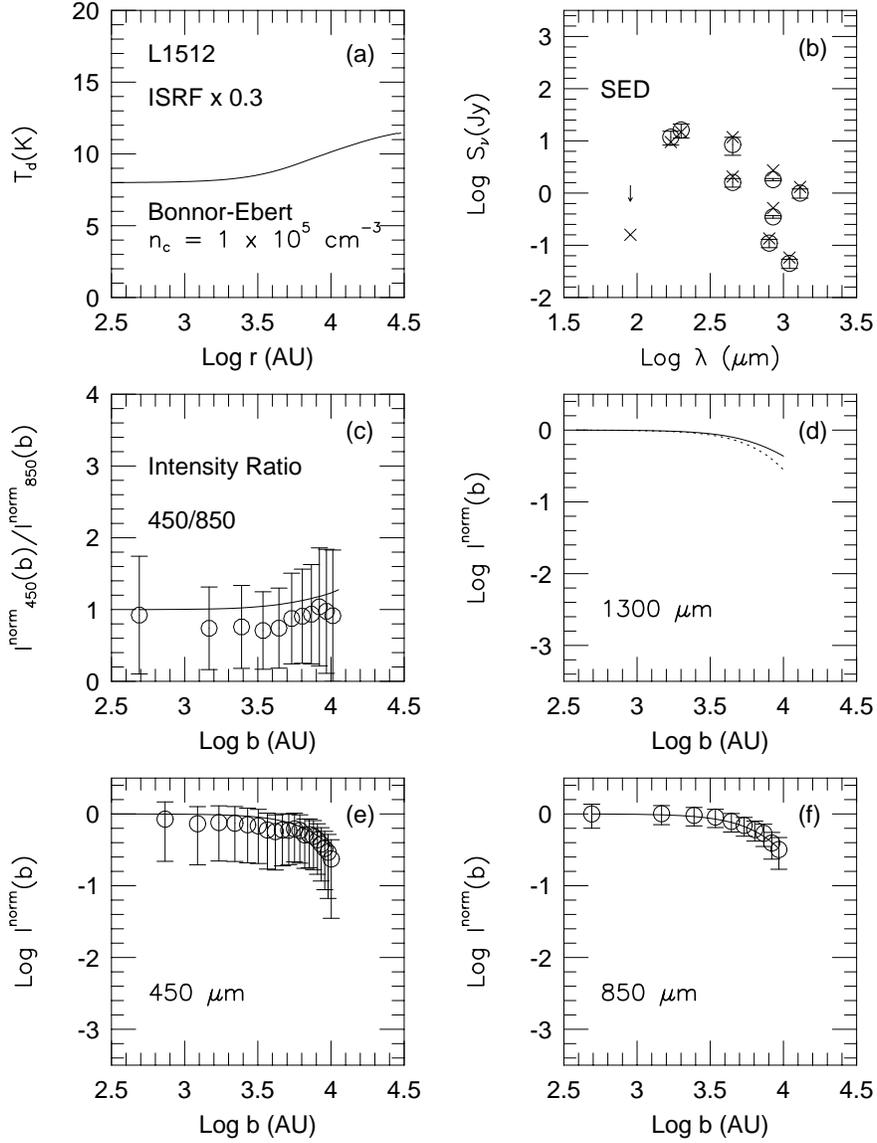}{6.5in}{0}{65}{65}{-200}{-20}
\caption{The same as Fig. \ref{figl1689b}, but for L1512.
The model is a Bonnor-Ebert sphere with $\nc = 1\ee5$ \cmv\ 
and a Black-Draine ISRF multiplied by 0.3.
No data exist at 1300 \micron, so only the model predictions are shown
(solid line for no chopping, dashed line for chopping by 120\arcsec).
At 140 pc, $b = 10^4$ AU corresponds to 71\arcsec.
 \label{figl1512}}
\end{figure}

\begin{figure}[hbt!]
\plotfiddle{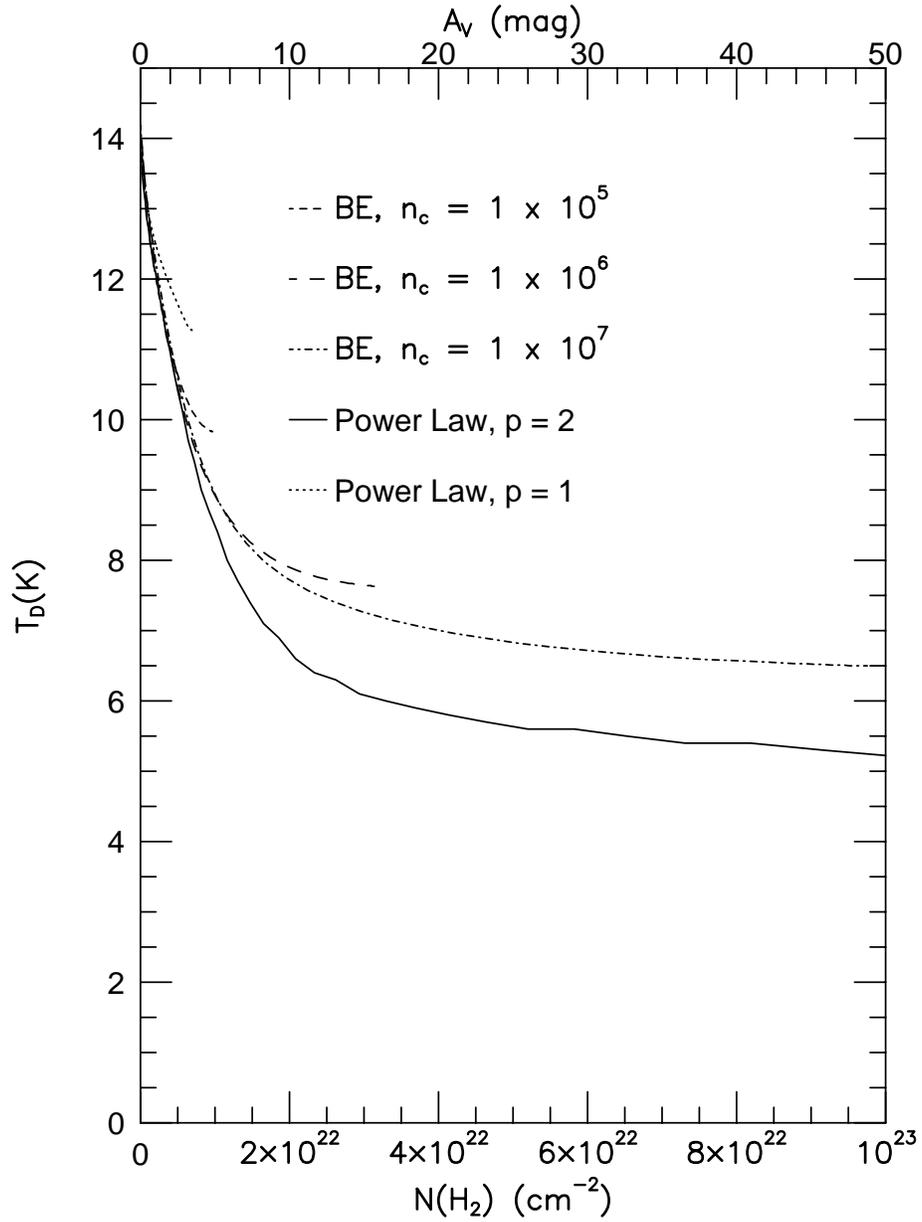}{6.5in}{0}{65}{65}{-200}{-20}
\caption{The dust temperature is plotted versus gas column density for
a variety of models. The visual extinction is marked on the top axis, using
standard conversions.  The calculation was done for the full Black-Draine
ISRF.
 \label{figcolden}}
\end{figure}

\begin{center} 
\begin{table}
\caption{Physical Models \label{physmods}}
\begin{tabular}{llrrrrrl}
\tableline\tableline
No. & Type\tablenotemark{a} & \router\tablenotemark{b} 
& $n_c$\tablenotemark{c} 
& $T_K$\tablenotemark{d} 
& $r_{flat}$\tablenotemark{e} 
& $n_{o}$\tablenotemark{f} 
& $M$\tablenotemark{g} 
\\
          &              & (AU)    
& (\cmv ) 
& (K)  
& (AU)  
& (\cmv)   
&  (\msun)      
\\
\tableline
1 & BE & 3\ee4 
& 1\ee4 
& 10 
& 16200 
& 1.84\ee3 
& 2.63 
\\
2 & BE & 3\ee4 
& 3\ee4 
& 10 
&  9300 
& 1.57\ee3 
& 3.15 
\\
3 & BE & 3\ee4 
& 1\ee5 
& 10 
&  5110 
& 1.17\ee3 
& 3.06 
\\
4 & BE & 3\ee4 
& 3\ee5 
& 10 
&  2990 
& 9.79\ee2 
& 2.75 
\\
5 & BE & 3\ee4 
& 1\ee6 
& 10 
&  1600 
& 9.38\ee2 
& 2.46 
\\
6 & BE & 3\ee4 
& 3\ee6 
& 10 
&   930 
& 9.90\ee2 
& 2.34 
\\
7 & BE & 3\ee4 
& 1\ee7 
& 10 
&   500 
& 1.09\ee3 
& 2.34 
\\
\tableline
8 & BE & 6\ee4 
& 3\ee5 
& 10 
&  2990 
& 2.34\ee2 
& 4.86 
\\
9 & BE & 1.5\ee4 
& 1\ee6 
& 10 
&1600 
& 3.98\ee3 
& 1.40 
\\
10 & BE & 6\ee4 
& 1\ee6 
& 10 
& 1600 
& 2.50\ee2 
& 4.64 
\\
11 & BE & 6\ee4 
& 3\ee6 
& 10 
&  930 
& 2.74\ee2 
& 4.69 
\\
\tableline
12 & BE & 3\ee4 
& 3\ee4 
& 5 
& 6590 
& 6.59\ee2 
& 1.58 
\\
13 & BE & 3\ee4 
& 1\ee5 
& 5 
& 3600 
& 5.14\ee2 
& 1.43 
\\
14 & BE & 3\ee4 
& 1\ee6 
& 5 
& 1130 
& 4.82\ee2 
& 1.18 
\\
15 & BE & 3\ee4 
& 1\ee6 
& 15 
& 1970 
& 1.41\ee3 
& 3.82 
\\
\tableline
16 & tBE$^4$ & 3\ee4 
& 3\ee5 
& $\sim$7-15 
& 2500 
& 1.68\ee3 
& 3.55 
\\
17 & tBE$^5$ & 3\ee4 
& 1\ee6 
& $\sim$7-15 
& 1300 
& 1.68\ee3 
& 3.42 
\\
18 & tBE$^{10}$ & 6\ee4 
& 1\ee6 
& $\sim$7-15 
& 1300 
& 4.27\ee2 
& 7.25 
\\
19 & tBE$^{7}$ & 3\ee4 
& 1\ee7 
& $\sim$7-15 
& 400  
& 1.77\ee3 
& 3.44 
\\
\tableline
20 & iPL2       & 3\ee4 
& 1.56\ee9  
& 10.4
& ... 
&  1.08\ee3 
& 2.06  
\\
21 & tPL2       & 3\ee4 
& 1.56\ee9  
& 10.4 
& ... 
&  1.08\ee3 
& 2.06  
\\
\end{tabular}
\tablenotetext{a}{BE is an isothermal Bonnor-Ebert sphere; tBE is a 
Bonnor-Ebert sphere with a temperature gradient; 
iPL2 is a power law, with an exponent of 2 and constant dust temperature of
10 K; tPL2
is the same power law with a dust temperature calculated by the radiative
transport code. }
\tablenotetext{b}{The outer radius is \router.}
\tablenotetext{c} {\nc\ is the central density, at $r = \rinner$.}
\tablenotetext{d} {\Tk\ is the gas kinetic temperature used to compute the
density distribution.}
\tablenotetext{e}{$r_{flat}$ is the radius where the density 
drops to half the central density (\nc).} 
\tablenotetext{f}{$n_o$ is the density at the outer radius.} 
\tablenotetext{g}{$M$ is the mass enclosed within \router.} 
\end{table}
\end{center}

\begin{center}
\begin{table}
\caption{Models of L1689B \label{l1689b.res}}
\begin{tabular}{rrrrrrrrl}
\tableline\tableline
No.\tablenotemark{a} & Type  & \nc\    & $M$  &  \chisq  & $\mean{\delta}$ &
\chisq\ & \chisq & Comments \\
    &       & (\cmv)    & (\msun) &  SCUBA   & SCUBA   & 
1300\tablenotemark{b}   & SED    &          \\
\tableline
3  & BE & 1\ee5 & 3.06  &  16 & 0.83 &  1036--1560 & 18 &     \\
4  & BE & 3\ee5 & 2.75  &  2.4 & 0.38 &  338--577  & 17 &     \\
5  & BE & 1\ee6 & 2.46  &  1.4 & 0.26 &  26--102 & 16 &     \\
6  & BE & 3\ee6 & 2.34  &  5.1 & 0.51 &  30--22 & 14 &     \\
7  & BE & 1\ee7 & 2.34  &  8.0 & 0.73 &  104--48 & 11 &     \\
\tableline
9  & BE & 1\ee6 & 1.40  &  1.2 & 0.26 & 27--58 &  6.6 &$\router = 1.5\ee4$  \\
8  & BE & 1\ee6 & 4.86  &  1.3 & 0.26 & 26--128 & 18  &$\router = 6\ee4$, $r_i = 50$  \\
5  & BE & 1\ee6 & 2.46  &  1.4 & 0.26 & 25--102 & 16 &$r_{i} = 50$    \\
\tableline
16 &tBE & 3\ee5 & 3.55  &  3.3 & 0.38 & 333--735 & 20 & \\
17 &tBE & 1\ee6 & 3.42  &  1.3 & 0.26 & 41--235 & 19 & \\
19 &tBE & 1\ee7 & 3.44  &  4.1 & 0.52 & 43--82 & 16 & \\
\tableline
4  & BE & 3\ee5 & 2.75  &  3.1 & 0.45 & 375--635 & 212 & OH2 Dust   \\
5  & BE & 1\ee6 & 2.46  &  1.0 & 0.27 & 39--131 & 237 & OH2 Dust     \\
\tableline
5  & BE & 1\ee6 & 2.46  &  1.4 & 0.26 & 26--102  & 14\tablenotemark{c}  & ISRF $\times 1$  \\
5  & BE & 1\ee6 & 2.46  &  1.1 & 0.27 & 31--123  & 1.5\tablenotemark{c} & ISRF $\times 0.5$  \\
\end{tabular}
\tablenotetext{a}{The model numbers refer to the physical model in Table 1.}
\tablenotetext{b}{The first value is for chopping by 120\arcsec; the
second is for no chopping.}
\tablenotetext{c}{Only these two models include the new \fir\ data in the
calculation of \chisq; the other models included previous \fir\ data.}
\end{table}
\end{center}

\begin{center}
\begin{table}
\caption{Models of L1544 \label{l1544.res}}
\begin{tabular}{rrrrrrrrl}
\tableline\tableline
No.\tablenotemark{a} & Type  & \nc\ & $M$ &  \chisq  & $\mean{\delta}$ 
& \chisq & \chisq & Comments \\
    &       &(\cmv) & (\msun) & SCUBA  & SCUBA   
& 1300\tablenotemark{b}   & SED    &          \\
\tableline
4  & BE   &  3\ee5  & 2.75  &   18   & 0.63  & 123--198 & 2.1 & \\
5  & BE   &  1\ee6  & 2.46  &   0.73 & 0.17  & 14--26 & 1.0 &  \\
6  & BE   &  3\ee6  & 2.34  &   3.0  & 0.19  & 118--49 & 1.2 &  \\
7  & BE   &  1\ee7  & 2.34  &   7.3  & 0.39  & 258--114 &  2.0 & \\
\tableline
17 & tBE  &  1\ee6  & 3.42  &   2.8  & 0.22  & 28--88 &  2.6 & \\
19 & tBE  &  1\ee7  & 3.44  &   2.0  & 0.24  & 160--88 &  2.0 & \\
\tableline
20 & iPL2 & \nodata & 2.06  &   22   & 0.95  &   480-247 &  11   \\
21 & tPL2 & \nodata & 2.06  &  3.0   & 0.27  &   149--210 & 3.6   \\
\tableline
5  & BE   & 1\ee6   & 2.46  &  0.73  & 0.17  & 14--26  & 7.2\tablenotemark{c} & IRSF $\times 1$   \\
5  & BE   & 1\ee6   & 2.46  &  1.1   & 0.22  & 13--31  & 2.8\tablenotemark{c} & IRSF $\times 0.6$   \\
\end{tabular}
\tablenotetext{a}{The model numbers refer to the physical model in Table 1.}
\tablenotetext{b}{The first value is for chopping by 120\arcsec; the
second is for no chopping.}
\tablenotetext{c}{Only these two models include the \fir\ data in the
calculation of \chisq.}
\end{table}
\end{center}

\begin{center}
\begin{table}
\caption{Models of L1512 \label{l1512.res}}
\begin{tabular}{rrrrrrrl}
\tableline\tableline
No.\tablenotemark{a}  & Type  & \nc   & $M$    & \chisq  
& $\mean{\delta}$ & \chisq & Comments \\
  &       &(\cmv) &(\msun) & SCUBA   & SCUBA           & SED    &          \\
\tableline
1   & BE & 1\ee4 & 2.63  & 3.2   & 0.93  & 6.6 \\
2   & BE & 3\ee4 & 3.15  & 1.1   & 0.62  & 62 \\
3   & BE & 1\ee5 & 3.06  & 0.38  & 0.41  & 179  \\
4   & BE & 3\ee5 & 2.75  & 1.6   & 0.59  & 276  \\
\tableline
12  & BE & 3\ee4 & 1.58  & 0.31  & 0.37  & 5.1 & $T_K = 5$ \\
13  & BE & 1\ee5 & 1.43  & 1.4   & 0.57  & 30 & $T_K = 5$ \\
14  & BE & 1\ee6 & 1.18  & 7.7   & 1.2   & 65 & $T_K = 5$ \\
15  & BE & 1\ee6 & 3.82  & 3.1   & 0.73  & 666 & $T_K = 15$ \\
\tableline
3   & BE & 1\ee5 & 3.06  & 0.38  & 0.41  & 155\tablenotemark{b} & ISRF $\times 1$ \\
3   & BE & 1\ee5 & 3.06  & 0.41  & 0.43  & 20\tablenotemark{b}  & ISRF $\times 0.3$ \\
\end{tabular}
\tablenotetext{a}{The model numbers refer to the physical model in Table 1.}
\tablenotetext{b}{Only these two models include the \fir\ data in the
calculation of \chisq.}
\end{table}
\end{center}

\end{document}